\documentclass[12pt]{article}
\usepackage{graphicx}
\usepackage{xspace}
\usepackage{multirow}
\usepackage{mciteplus}

\usepackage{ifthen} % for conditional statements
\newboolean{uprightparticles}
\setboolean{uprightparticles}{false} %True for upright particle symbols
\newboolean{articletitles}
\setboolean{articletitles}{true} % False removes titles in references
\newboolean{inbibliography}
\setboolean{inbibliography}{false} %True once you enter the bibliography

\def\MagUp {\mbox{\em Mag\kern -0.05em Up}\xspace}

\ifthenelse{\boolean{uprightparticles}}%
{

 \def\Ppi         {\ensuremath{\uppi}\xspace}

 \def\PDelta      {\ensuremath{\Delta}\xspace}                 
 \def\PXi      {\ensuremath{\Xi}\xspace}                 
 \def\PLambda      {\ensuremath{\Lambda}\xspace}                 
 \def\PSigma      {\ensuremath{\Sigma}\xspace}                 
 \def\POmega      {\ensuremath{\Omega}\xspace}                 
 \def\PUpsilon      {\ensuremath{\Upsilon}\xspace}                 
 
 \def\PB      {\ensuremath{\mathrm{B}}\xspace}                 
                  
 \def\PD      {\ensuremath{\mathrm{D}}\xspace}

 \def\PK      {\ensuremath{\mathrm{K}}\xspace}

 \def\Pi      {\ensuremath{\mathrm{i}}\xspace}

 \def\Ps      {\ensuremath{\mathrm{s}}\xspace}

}
{

 \def\Ppi         {\ensuremath{\pi}\xspace}

 \mathchardef\PDelta="7101
 \mathchardef\PXi="7104
 \mathchardef\PLambda="7103
 \mathchardef\PSigma="7106
 \mathchardef\POmega="710A
 \mathchardef\PUpsilon="7107
                  
 \def\PB      {\ensuremath{B}\xspace}                 
                  
 \def\PD      {\ensuremath{D}\xspace}

 \def\PK      {\ensuremath{K}\xspace}

 \def\Pi      {\ensuremath{i}\xspace}

 \def\Ps      {\ensuremath{s}\xspace}

}

\makeatletter
\ifcase \@ptsize \relax% 10pt
  \newcommand{\miniscule}{\@setfontsize\miniscule{4}{5}}% \tiny: 5/6
\or% 11pt
  \newcommand{\miniscule}{\@setfontsize\miniscule{5}{6}}% \tiny: 6/7
\or% 12pt
  \newcommand{\miniscule}{\@setfontsize\miniscule{5}{6}}% \tiny: 6/7
\fi
\makeatother

\DeclareRobustCommand{\optbar}[1]{\shortstack{{\miniscule (\rule[.5ex]{1.25em}{.18mm})}
  \\ [-.7ex] $#1$}}

\def\squark    {{\ensuremath{\Ps}}\xspace}

\def\pion   {{\ensuremath{\Ppi}}\xspace}

\def\pip    {{\ensuremath{\pion^+}}\xspace}
\def\pim    {{\ensuremath{\pion^-}}\xspace}

\def\kaon    {{\ensuremath{\PK}}\xspace}
  \def\Kbar    {{\kern 0.2em\overline{\kern -0.2em \PK}{}}\xspace}

\def\KorKbar    {\kern 0.18em\optbar{\kern -0.18em K}{}\xspace}
\def\Kz      {{\ensuremath{\kaon^0}}\xspace}

\def\Kp      {{\ensuremath{\kaon^+}}\xspace}
\def\Km      {{\ensuremath{\kaon^-}}\xspace}

\def\KS      {{\ensuremath{\kaon^0_{\rm\scriptscriptstyle S}}}\xspace}

  \def\Dbar    {{\kern 0.2em\overline{\kern -0.2em \PD}{}}\xspace}

\def\DorDbar    {\kern 0.18em\optbar{\kern -0.18em D}{}\xspace}

\def\Dzb     {{\ensuremath{\Dbar{}^0}}\xspace}

\def\B       {{\ensuremath{\PB}}\xspace}
\def\Bbar    {{\ensuremath{\kern 0.18em\overline{\kern -0.18em \PB}{}}}\xspace}

\def\BorBbar    {\kern 0.18em\optbar{\kern -0.18em B}{}\xspace}
\def\Bz      {{\ensuremath{\B^0}}\xspace}

\def\Bs      {{\ensuremath{\B^0_\squark}}\xspace}

  \def\Y#1S{\ensuremath{\PUpsilon{(#1S)}}\xspace}% no space before {...}!

\def\Lbar        {{\ensuremath{\kern 0.1em\overline{\kern -0.1em\PLambda}}}\xspace}
\def\LorLbar    {\kern 0.18em\optbar{\kern -0.18em \PLambda}{}\xspace}

\def\to                 {\ensuremath{\rightarrow}\xspace}

\def\AT#1     {\ensuremath{A_{\mathrm{T}}^{#1}}\xspace}           % 2

\def\C#1      {\ensuremath{\mathcal{C}_{#1}}\xspace}                       % 9
\def\Cp#1     {\ensuremath{\mathcal{C}_{#1}^{'}}\xspace}                    % 7
\def\Ceff#1   {\ensuremath{\mathcal{C}_{#1}^{\mathrm{(eff)}}}\xspace}        % 9  
\def\Cpeff#1  {\ensuremath{\mathcal{C}_{#1}^{'\mathrm{(eff)}}}\xspace}       % 7
\def\Ope#1    {\ensuremath{\mathcal{O}_{#1}}\xspace}                       % 2
\def\Opep#1   {\ensuremath{\mathcal{O}_{#1}^{'}}\xspace}                    % 7

\newcommand{\tev}{\ifthenelse{\boolean{inbibliography}}{\ensuremath{~T\kern -0.05em eV}\xspace}{\ensuremath{\mathrm{\,Te\kern -0.1em V}}}\xspace}
\newcommand{\gev}{\ensuremath{\mathrm{\,Ge\kern -0.1em V}}\xspace}
\newcommand{\mev}{\ensuremath{\mathrm{\,Me\kern -0.1em V}}\xspace}
\newcommand{\kev}{\ensuremath{\mathrm{\,ke\kern -0.1em V}}\xspace}
\newcommand{\ev}{\ensuremath{\mathrm{\,e\kern -0.1em V}}\xspace}
\newcommand{\gevc}{\ensuremath{{\mathrm{\,Ge\kern -0.1em V\!/}c}}\xspace}
\newcommand{\mevc}{\ensuremath{{\mathrm{\,Me\kern -0.1em V\!/}c}}\xspace}
\newcommand{\gevcc}{\ensuremath{{\mathrm{\,Ge\kern -0.1em V\!/}c^2}}\xspace}
\newcommand{\gevgevcccc}{\ensuremath{{\mathrm{\,Ge\kern -0.1em V^2\!/}c^4}}\xspace}
\newcommand{\mevcc}{\ensuremath{{\mathrm{\,Me\kern -0.1em V\!/}c^2}}\xspace}

\def\invfb   {\ensuremath{\mbox{\,fb}^{-1}}\xspace}

\def\invab   {\ensuremath{\mbox{\,ab}^{-1}}\xspace}

\def\gsim{{~\raise.15em\hbox{$>$}\kern-.85em
          \lower.35em\hbox{$\sim$}~}\xspace}
\def\lsim{{~\raise.15em\hbox{$<$}\kern-.85em
          \lower.35em\hbox{$\sim$}~}\xspace}

\def\tell1  {TELL1\xspace}
\def\ukl1   {UKL1\xspace}

\newcommand\pubnumber{ } % left blank for now
\newcommand\pubdate{November 17, 2014}

\def\lpnhe{LPNHE, Universit\'{e} Pierre et Marie Curie\\
Universit\'{e} Paris Diderot, CNRS/IN2P3, Paris, France}
\def\support{\footnote{Work supported by CNRS/IN2P3}}

\def\Title#1{\begin{center} {\Large #1 } \end{center}}
\def\Author#1{\begin{center}{ \sc #1} \end{center}}
\def\Address#1{\begin{center}{ \it #1} \end{center}}

\newcommand\pubblock{\rightline{\begin{tabular}{l} \pubnumber\\
         \pubdate  \end{tabular}}}
\newenvironment{Abstract}{\begin{quotation}  }{\end{quotation}}
\newenvironment{Presented}{\begin{quotation} \begin{center} 
             PRESENTED AT\end{center}\bigskip 
      \begin{center}\begin{large}}{\end{large}\end{center} \end{quotation}}

\def\beq{\begin{equation}}
\def\eeq#1{\label{#1}\end{equation}}
\def\eeqn{\end{equation}}

\def\beqa{\begin{eqnarray}}
\def\eeqa#1{\label{#1}\end{eqnarray}}
\def\eeqan{\end{eqnarray}}

\let\bar=\overbar

\def\Dslash{\not{\hbox{\kern-4pt $D$}}}
\def\dslash{\not{\hbox{\kern-2pt $\del$}}}

\def\msb{{\bar{\ssstyle M \kern -1pt S}}}

\begin{document}
\begin{titlepage}
\pubblock

\vfill
\Title{CKM studies in the charm sector}
\vfill
\Author{M.J. Charles\support}
\Address{\lpnhe}
\vfill
\begin{Abstract}
A summary of recent progress in charm mixing and CP violation
is presented, with a heavy bias towards experimental results.
\end{Abstract}
\vfill
\begin{Presented}
The 8th International Workshop on the CKM Unitarity Triangle (CKM 2014),\\
Vienna, Austria, September 8--12, 2014
\end{Presented}
\vfill
\end{titlepage}
\def\thefootnote{\fnsymbol{footnote}}
\setcounter{footnote}{0}

\section{Introduction}
\label{sec:intro}

Famously, CKM studies of charm began even before its experimental
discovery~\cite{Glashow:1970gm}.
The experimental and theoretical context has changed a great deal
since then, but the goal today remains the same: to reveal or
constrain New Physics (NP) through study of flavour observables.
For charm physics this means making precise measurements of
CP violation (CPV) and mixing\footnote{
  The branching fractions of very rare decays such as
  $D^0 \to \mu^+ \mu^-$ are also sensitive to NP,
  but are outside the scope of these Proceedings.
}. CPV in charm is typically divided into two types
according to its mechanism:
indirect (mixing-induced) and direct.
Although both are ultimately driven by CKM physics---at least,
within the Standard Model (SM)---they are studied in
different ways at both theoretical and experimental levels,
and will be discussed separately.

\section{Mixing}
\label{sec:mixing}

Charm is far from the only arena where flavour observables can be
used to search for NP. In particular, numerous studies of mixing
and CPV have been carried out in the $b$ and $s$ sectors.
The differences are stark. Mixing occurs at a much higher rate
in \Kz, \Bz, and \Bs mesons, as shown in
Table~\ref{tab:mixing}. The normalised mixing parameters quoted,
$x$ and $y$, are defined via
\begin{eqnarray*}
  \Gamma &=& \frac{\Gamma_2 + \Gamma_1}{2} \\
  x &=& \frac{m_2 - m_1}{\Gamma} \\
  y &=& \frac{\Gamma_2 - \Gamma_1}{\Gamma}
\end{eqnarray*}
where $m_i$ and $\Gamma_i$ are the masses and widths of the
eigenstates of the Hamiltonian. The overall mixing rate $R_M$
is then given by $R_M = (x^2 + y^2)/2$. For each of the mesons
able to mix, this rate is of order 1 or more---except for
charm, where it is approximately $3 \times 10^{-5}$.
This small rate is why it took about three decades from the
discovery of charm to the first evidence of
mixing~\cite{Aubert:2007wf,Staric:2007dt}.
Even so, the mixing parameters are at the upper end of what
is plausible within the SM: long-distance effects can generate
values of $x$ and $y$ up to
$\mathcal{O}(10^{-2})$~\cite{Falk:2004wg,Petrov:2013usa}.
Short-distance effects would be much smaller,
with a simple estimate indicating they should enter
at the $10^{-5}$ level~\cite{Bianco:2003vb};
more sophisticated calculations give numbers that
are larger but not dramatically so~\cite{Bobrowski:2012jf}.
It should be noted that all calculations of the SM mixing
rate to date are at best order-of-magnitude: for all we
know, the mixing rates observed could actually be
dominated by NP effects. Inverting the argument, the
observed level of mixing in charm can be used to
constrain models of NP by requiring that they at most
saturate it~\cite{Golowich:2007ka}.

\begin{table}
  \begin{center}
    \begin{tabular}{ccc}  
      Meson & $x$ & $y$ \\ \hline
      \Kz & $9.5$  & almost 1 \\
      $D^0$ & $0.0041 \pm 0.0015$ & $0.0063 \pm 0.0008$  \\
      \Bz & $0.774 \pm 0.006$ & $0.0005 \pm 0.005$ \\
      \Bs & $26.85 \pm 0.13$ & $0.069 \pm 0.006$
    \end{tabular}
    \caption{Mixing parameters $x$ and $y$ for neutral mesons.
      Data from HFAG~\cite{bib:hfag}
      and the PDG~\cite{bib:pdg}.
    }
    \label{tab:mixing}
  \end{center}
\end{table}

On the experimental side, there are four main sources of
information on charm mixing\footnote{
  This list is not exhaustive. For example, it omits
  time-integrated measurements of wrong-sign semileptonic
  decays.
}:
the time-dependent ratio of wrong-sign (WS) to right-sign (RS)
  decays, primarily in $D^0 \to K^{\pm} \pi^{\mp}$, measuring $y_{CP}$;
the ratio of the mean lifetime seen in CP eigenstates vs.\ CP-mixed
  states, primarily $D^0 \to \Km\Kp, \pim\pip$ vs.\ $\Km \pip$, measuring $x^{\prime}$ and $y^{\prime}$\footnote{
    $x^{\prime}$ and $y^{\prime}$ are related to $x$ and $y$ by rotation through the strong phase $\delta$
    between the DCS and CF decays to that final state.
};
time-dependent amplitude analyses of Dalitz plots of self-conjugate final states, primarily
  $D^0 \to \KS \pip \pim$, measuring $x$ and $y$; and
measurements of relative strong phases $\delta$ between 
  doubly Cabibbo-suppressed (DCS) and Cabibbo-favoured (CF) decays.
In most cases these do not grant direct access to $(x,y)$ but
rather combinations of physics parameters ($x$, $y$, relative
strong phases, indirect CPV), such that the world-average is not
dominated by any one type of measurement but rather by their combination.
The time-dependent amplitude analyses are the exception:
they allow simultaneous measurement of the mixing rate and of the
strong phases of the contributing amplitudes, thereby granting direct
access to $x$ and $y$.

\begin{figure}
  \begin{center}
    \begin{tabular}{cc}
      \includegraphics[width=0.49\textwidth]{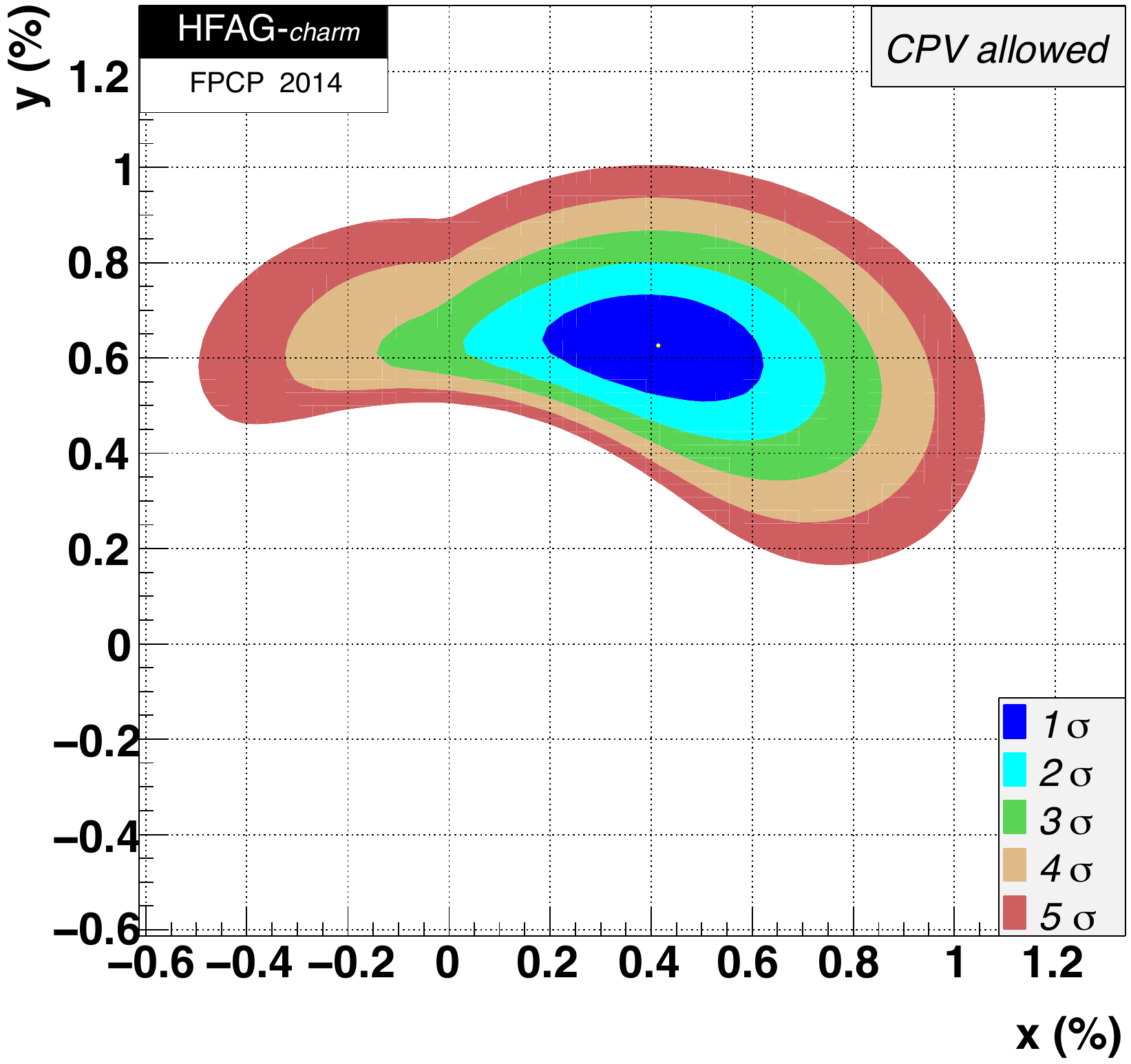} &
      \includegraphics[width=0.49\textwidth]{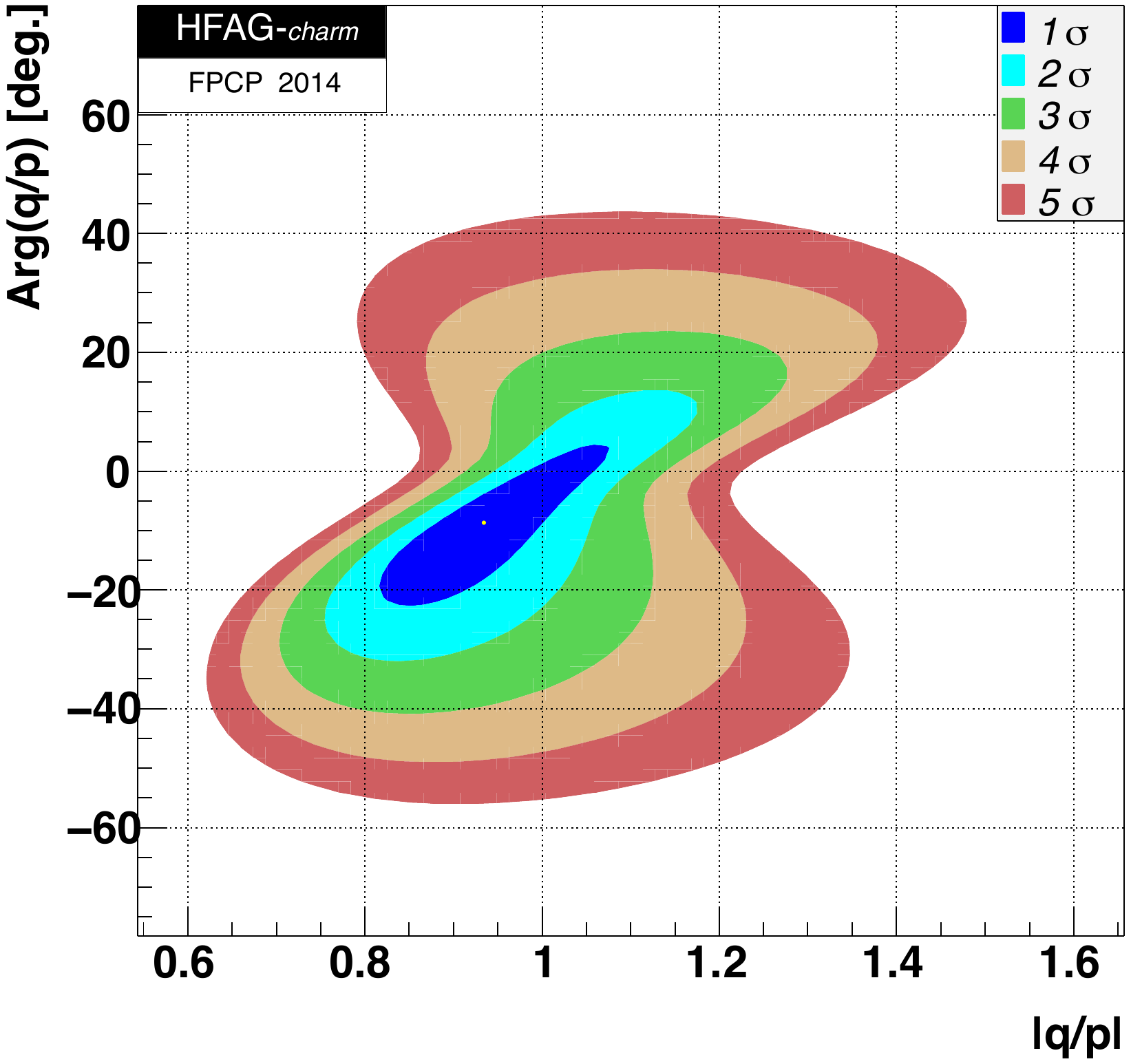} \\ (a) & (b)
    \end{tabular}
  \end{center}
  \caption{World-average results from HFAG~\cite{bib:hfag}, as of 30 June 2014.}
  \label{fig:hfag}
\end{figure}

The most recent world average results on $x$ and $y$ from HFAG
are illustrated in Fig.~\ref{fig:hfag}(a). Allowing for CPV, the central values are:
\begin{eqnarray*}
  x &=& \left( 0.41 ^{+0.14}_{-0.15} \right) \% \\
  y &=& \left( 0.63 ^{+0.07}_{-0.08} \right) \%
\end{eqnarray*}
This represents a major improvement in precision from the status as of CKM~2012.
Several new or updated experimental results have been released since then, from 
Belle~\cite{Peng:2014oda,Ko:2014qvu},
CDF~\cite{Aaltonen:2013pja}
LHCb~\cite{Aaij:2012nva,Aaij:2013wda},
and BESIII~\cite{Ablikim:2014gvw,Xiao-RuiLufortheBESIII:2013qna}.
In particular, after many years in which the overall statistical significance
of mixing was high but no individual measurement exceeded $5\sigma$,
Belle, CDF, and LHCb have now all passed that threshold.
The most significant of these results by some way is
a measurement with WS $D^0 \to K^+ \pi^-$ decays by LHCb:
this is an analysis which can capitalise on the large charm
cross-section and the boost.

As discussed above, the size of mixing in charm is a powerful constraint on
models of NP. However, there is not much more to be learned on this front
from experiment: improving the precision on the mixing parameters will not
move the upper bound significantly. It is important for another reason,
though: relating measurements of time-dependent CP asymmetries
to the underlying physics parameters of indirect CP violation
requires knowledge of $x$ and $y$.

\section{Indirect CPV}
\label{sec:indirect}

Indirect CPV in charm may be characterised by the magnitude and
phase of $q/p$, with deviation from $|q/p|=1$ and
$\phi \equiv \mathrm{arg}(q/p) = 0$ indicating CPV.
In the HFAG phase convention~\cite{bib:hfag},
these are related to experimental observables as follows:\footnote{
  Direct CPV is neglected in these expressions.
  This is a good approximation at present, but when the precision
  on $A_{\Gamma}$ improves it may need to be taken into
  account~\cite{Gersabeck:2011xj}.
}
\begin{eqnarray*}
  2 y_{CP} = \left( |q/p| + |p/q| \right) y \cos \phi -  \left( |q/p| - |p/q| \right) x \sin \phi \\
  2 A_{\Gamma} = \left( |q/p| - |p/q| \right) y \cos \phi -  \left( |q/p| + |p/q| \right) x \sin \phi \\
  x^{\prime \pm} = \left( \frac{1 \pm A_M}{1 \pm A_M} \right)^{1/4} \, \left( x^{\prime} \cos \phi \pm y^{\prime} \sin \phi \right) \\
  y^{\prime \pm} = \left( \frac{1 \pm A_M}{1 \pm A_M} \right)^{1/4} \, \left( y^{\prime} \cos \phi \mp x^{\prime} \sin \phi \right) \\
\end{eqnarray*}
where
\begin{eqnarray*}
  A_M = \frac{ |q/p|^2 - |p/q|^2 }{ |q/p|^2 + |p/q|^2 }
  .
\end{eqnarray*}
Here $y_{CP}$ is the mixing parameter obtained from the ratio of
CP-even\footnote{
  CP-odd final states can also be used, with suitable
  changes to the formulae.
} to CP-mixed lifetimes, discussed in the previous section.
$A_{\Gamma}$ is a related asymmetry:
\begin{eqnarray*}
  A_{\Gamma} = \frac{ \tau(\Dzb \to f_{CP,+}) - \tau(D^0 \to f_{CP,+}) }{ \tau(\Dzb \to f_{CP,+}) + \tau(D^0 \to f_{CP,+}) }
  ,
\end{eqnarray*}
with the most precise results coming from
$D^0 \to \Kp\Km$ and $\pip\pim$.
The observables $x^{\prime \pm}$ and $y^{\prime \pm}$ are simply
the values of $x^{\prime}$ and $y^{\prime}$ measured with
$D^0$ and $\Dzb$ events, respectively; a difference
between $x^{\prime+}$ and $x^{\prime-}$ or $y^{\prime+}$ and $y^{\prime-}$
would indicate CPV.

Two key points follow directly from these expressions.
First, the observable $A_{\Gamma}$ is scaled down by a factor
of $x$ or $y$, such that even if indirect CPV is maximal
$A_{\Gamma}$ will be limited in magnitude to approximately $x$ or $y$,
i.e.\ $\mathcal{O}(\frac{1}{2}\%)$.
We have only recently entered a regime where the uncertainty on
$A_{\Gamma}$ is smaller than this.
Second, while a non-zero value of $A_{\Gamma}$ would necessarily
imply CPV, extracting values or limits on the physics parameters
$\phi$ and $|q/p|$ would require $x$ and $y$ as inputs.
Similar considerations apply to $x^{\prime \pm}$ and $y^{\prime \pm}$. 
The relative uncertainties on the mixing parameters are
still substantial, although the situation is much better now than
a few years ago. Thus, improving the precision with which $x$ and $y$
is known translates directly to improved constraints on the underlying
CPV parameters. 

Some of the recent mixing measurements noted in Section~\ref{sec:mixing}
included measurements of CP asymmetries~\cite{Peng:2014oda,Aaij:2013wda},
and in addition LHCb has published a measurement of $A_{\Gamma}$ with
1/invfb of data~\cite{Aaij:2013ria}.
Including these, the current world-average results with CPV fully allowed
are shown in Fig.~\ref{fig:hfag}(b), and correspond to:
\begin{eqnarray*}
  |q/p| &=& 0.93 ^{+0.09}_{-0.08} \\
  \phi  &=& \left( -8.7 ^{+8.7}_{-9.1} \right)^{\circ}
  .
\end{eqnarray*}
Those values allow for the presence of direct CPV in suppressed decays.
If it is assumed that there is no direct CPV in DCS decays\footnote{
  This is certainly true for charm in the SM, and is also true in many but not 
  all models of NP on the market. However, we should be a little cautious
  simply because there have been few experimental tests of this.
}~\cite{Kagan:2009gb}, these constraints are
strengthened considerably:
\begin{eqnarray*}
  |q/p| &=& 1.007 ^{+0.015}_{-0.014} \\
  \phi  &=& (-0.30 ^{+0.58}_{-0.60})^{\circ}
  .
\end{eqnarray*}
As with the mixing parameters, these represent a great improvement on
the state of play as of CKM~2012. It is now clear that we do not have
$\mathcal{O}(1)$ indirect CPV in charm. If direct CPV is negligible in
DCS decays---and the excellent agreement with the null hypothesis
seems to back this up---then $\mathcal{O}(10^{-1})$ is also excluded
and even $\mathcal{O}(10^{-2})$ is not far away. We have not yet
reached the SM level, which is estimated to be
$\mathcal{O}(10^{-4})$ to $\mathcal{O}(10^{-3})$~\cite{Bigi:2009aw,Bobrowski:2010xg,Bobrowski:2013vak},
but it looms ahead in the distance. On the plus side, it is
dominated by short-distance physics so is a more theoretically
tractable problem than direct CP violation.

\section{Direct CPV}

While indirect CPV is not quite universal in the SM---CPV in the interference
between mixing and decay in general depends on the final state---it can
be expressed in terms of a small number of parameters with modest assumptions.
The story is very different for direct CPV, where the asymmetry can
vary greatly from one decay mode to another.
Both within the SM and in general models of NP, the largest asymmetries are
expected in SCS modes~\cite{Grossman:2006jg} where interference between
penguin and tree diagrams can be substantial, and this is where both
experimental and theoretical interest has been focused.
Direct CPV in DCS charm decays is possible in certain models of NP but negligible
in the SM; the experimental precision is necessarily poorer than for SCS decays
simply because of the reduced branching fractions.
In CF decays, direct CPV is negligible even in the presence of NP since the
favoured tree diagram dominates.

There was considerable excitement at the time of CKM~2012:
LHCb had seen indications of direct CPV at $3.5\sigma$ in
$\Delta A_{CP} = A_{CP}(D^0 \to \Kp\Km) - A_{CP}(D^0 \to \pip\pim)$
with 0.6\invfb of prompt charm data~\cite{Aaij:2011in}.
CDF and Belle reported similar central values after analysing their full
data sets~\cite{Collaboration:2012qw,Ko:2012px}, giving a world average
of $\Delta A_{CP}^{\mathrm{dir}} = (-0.678 \pm 0.147)\%$.
The effect was studied in numerous theory papers
(e.g. \cite{Brod:2012ud,Feldmann:2012js,Bhattacharya:2012ah,Isidori:2011qw}),
both in the context of the SM and of various models of NP.
The consensus was that direct CPV of order $0.5\%$ in these final states could
be accommodated within the SM if the penguin amplitude is large, which is
allowed and would be consistent with the unusual pattern of branching fractions
[$\mathcal{B}(D^0 \to \Kp\Km) > \mathcal{B}(D^0 \to \pip\pim)$].
Since CKM~2012, LHCb has released three new results which have brought the
world average much closer to zero:
an update of the prompt charm measurement to 1\invfb~\cite{LHCb-CONF-2013-003},
a measurement with charm produced in semileptonic $B$ decays with 1\invfb~\cite{Aaij:2013bra},
and an update of the latter measurement to 3\invfb~\cite{Aaij:2014gsa}.
With these new results, the HFAG average~\cite{bib:hfag} is
$\Delta A_{CP}^{\mathrm{dir}} = (-0.253 \pm 0.104)\%$,
consistent with zero CPV.

As well as updating the $\Delta A_{CP}$ measurement, the 3\invfb LHCb paper
mentioned above~\cite{Aaij:2014gsa} also quoted values for the individual asymmetries
$A_{CP}(D^0 \to \Kp\Km)$ and $A_{CP}(D^0 \to \pip\pim)$.
Because the $pp$ initial state at LHCb is not charge antisymmetric (in the way that
an $e^+ e^-$ or $p \bar{p}$ collision is), measuring these asymmetries requires
correcting for production and efficiency asymmetries. These were obtained from
a careful combination of control modes, extending a technique that had been
used previously at LHCb~\cite{Aaij:2013ula,Aaij:2013wda}.
This is a promising development for future measurements at LHCb.

Numerous other results on CPV have been reported or published since CKM~2012, from
CLEO~\cite{Bonvicini:2013vxi},
BaBar~\cite{Lees:2012jv,Lees:2012nn},
Belle~\cite{Nisar:2014fkc,Ko:2012uh}
and LHCb~\cite{Aaij:2013ula,Aaij:2014qec,Aaij:2013swa,Aaij:2013jxa,Aaij:2014qwa}.
None report a significant CP asymmetry.
While there is not space here to discuss all of these measurements,
there are a couple of points worth noting.
One is that the analysis technology used to study CP asymmetries
in multi-body is advancing rapidly. Model-independent techiques were
pioneered in 3-body decays with studies of binned Dalitz
plots~\cite{Aubert:2008yd,Bediaga:2009tr}
and have proven their sensitivity in $B$ decay modes
to similar final states~\cite{Aaij:2014iva}.
However, there is always a question as to how to choose the binning:
the optimal choice depends on the model of CP violation.
Unbinned methods~\cite{Williams:2010vh} are now being applied.
These are especially interesting for decays to four-body (or more!)
final states, where a tradiational binned analysis (e.g.~\cite{Aaij:2013swa})
runs swiftly into the curse of dimensionality.
Another item worth singling out is the new Belle result on
$A(D^0 \to \pi^0 \pi^0)$~\cite{Nisar:2014fkc}. This is
an analysis that LHCb cannot hope to do, and is an
excellent example of the complementarity between hadron
and $e^+ e^-$ experiments that we will see in the
Belle-II/LHCb upgrade era.

All of this experimental progress is very good news:
multiple two-body decay modes reaching precisions well below 1\%
and closing in on $10^{-3}$, and searches for CPV in the phase
space of multi-body decays with signal yields in the millions.
The theoretical side, however, remains unclear:
one key lesson from $\Delta A_{CP}$ is that direct CP asymmetries
can arise at the $\mathcal{O}(10^{-3})$ level in the SM,
and as much as $\mathcal{O}(\mathrm{few} \times 10^{-3})$ can be
accommodated.
We have already reached this level of precision, so in the
absence of a major advance in theoretical engineering it
seems like we would not be able to say whether some future
observation of direct CPV in a generic SCS mode at the
$0.1\%$-level is SM physics or not.
A path forward has been suggested: to combine
information from $SU(3)$-related final states,
testing whether the overall pattern is
consistent with the SM~\cite{Hiller:2012xm}.
Alternatively, effort can be focused on modes where the
SM contribution is suppressed
(e.g. SCS modes with $\Delta I = 3/2$~\cite{Bhattacharya:2012ah}).

One last note: charmed baryons are also produced at significant rates
at LHCb~\cite{Aaij:2013mga}, and good-sized samples have been
accumulated at the $B$-factories. Their decays are also sensitive
to direct CPV, but measurements of this are sorely lacking.

\section{The future}

From an experimental point of view, the coming years will
be a very exciting time for charm physics.
At the time of writing, the LHC is near the end of a long
shutdown during which the machine is upgraded to operate
at $\sqrt{s} \approx 13$\,TeV. Data-taking is scheduled
to resume in 2015, with Run~2 lasting around three
years. During this time LHCb will integrate
about 5--6\invfb. This will give the charm statistics
a major boost: the new data should correspond to about
twice the current integrated luminosity, at a higher
energy (i.e. nearly twice the production cross-section),
with an improved software trigger. Indirect CPV measurements
in $D^0 \to h^+ h^{\prime -}$ with this sample will
have an exciting level of sensitivity to NP.
In parallel, Belle~II will begin commissioning in 2015
and should start to integrate luminosity in earnest
around 2018. Another LHC shutdown is planned around
2018--2019, during which the LHCb detector will be
upgraded to allow a factor 5--10 increase in
instantaneous luminosity. In total, LHCb intends
to integrate about 50\invfb, and Belle II 50\invab.
(Note that the cross-sections at the two colliders are
different, such that the LHCb sample corresponds to
more $c \bar{c}$ pairs produced.)

Both experiments have published estimates of their
long-run physics sensitivity~\cite{Aushev:2010bq,Bediaga:2012py}.
While such estimates should be treated with a degree
of caution\footnote{
  This in a spirit of optimism: ``more data makes us smarter''.
}, we can expect an improvement of at least an order of
magnitude in the uncertainties on charm CP asymmetries.
There is good complementarity between the two experiments:
the sheer statistics at LHCb will pin down key observables
in decays to $D^0 \to h^+h^{\prime-}$ and
$D_{(s)}^+ \to h^+ h^{\prime+} h^{\prime\prime-}$ final states;
Belle II will give us depth, opening up many modes that
LHCb struggles with;
and redundancy between the two will validate our control
over systematic uncertainties.

Last but not least, we can look forward to 
new results from charm-at-threshold experiments
such as BESIII.
These are one of the only sources of information
on the strong phases $\delta$ that feed into the
mixing average discussed in Section~\ref{sec:mixing}.
In particular, separate measurements of the strong phase in
Dalitz plot bins for $D^0 \to \KS h^+ h^-$
allows for model-independent measurements of
$x$ and $y$~\cite{Libby:2010nu,Thomas:2012qf}
at LHCb and Belle~II.

\section{Measurements reported during CKM~2014}

These proceedings are based on a talk given in the opening session.
Several new experimental results, listed below, were reported during the conference.
For more details please consult the proceedings from those talks.
\begin{itemize}
  \item Measurements of $A_{\Gamma}$ in
    $D^0 \to \Kp\Km$, $\pip\pim$ at CDF~\cite{Aaltonen:2014efa}.
    Combining the two modes, an average value of
    $A_{\Gamma} = (-0.12 \pm 0.12)\%$ was reported.
    Presented by L.~Sabato.
  \item A time-integrated search for CPV in
    $D^0 \to \pi^+ \pi^- \pi^0$ at LHCb~\cite{Aaij:2014afa}.
    An unbinned, model-independent method was used with a
    sample of approximately 0.6M signal.
    No evidence for CPV was reported, with a $p$-value for
    consistency with the null hypothesis of $(2.6 \pm 0.5)\%$.
    Presented by E.~Gersabeck.
  \item Preliminary results from BaBar on $P$, $C$, and $CP$
    violation in 4-body $D$ decays, following a method
    proposed recently~\cite{Bevan:2014nva}
    to reinterpret existing measurements of $T$-odd correlations.
    Presented by M.~Martinelli.
\end{itemize}

\section{Conclusions}

There has been an enormous amount of progress in the past two years,
with new results from several experiments. After a ramp-up, the
LHCb charm machine is now in high gear, producing a steady stream
of measurements. We can look forward to a similar flow from Belle II
in the not-too-distant future.

After a burst of excitement, $\Delta A_{CP}$ now looks firmly consistent
with the Standard Model once again. The same is true across the board
in charm: we find no evidence of anything amiss. The game is far from
over, however: we can expect a jump in precision for indirect CPV in
the next few years, followed by steady progress as the LHCb upgrade
takes data. We are still operating in a regime where the SM is---at least
in principle---safely below current sensitivity to indirect CPV,
so that any observation would point to NP. For direct CPV the situation
is less clear, so all we can do is plough ahead, measure as much as
we can in as many modes as we can, and try to make sense of whatever
nature decides to give us.

\setboolean{inbibliography}{true}
\bibliographystyle{LHCb}
\bibliography{proc,LHCb-CONF}

\end{document}